\documentclass[shortnote,twocolumn]{jpsj3}
\usepackage{color}
\usepackage{txfonts}
\usepackage{bm}
\usepackage{mhchem}

\DeclareMathOperator{\diag}{diag}

\title{Note on Angular Momentum of Phonons in Chiral Crystals}
\author{Akihito Kato$^1$ and Jun-ichiro Kishine$^{2,3}$}
\inst{$^1$Department of Physics and Electronics, Osaka Metropolitan University, 1-1 Naka-ku, Sakai, Osaka 599-8531, Japan \\
$^2$Division of Natural and Environmental Sciences, The Open University of Japan, Chiba 261-8586, Japan
\\
$^3$Quantum Research Center for Chirality, Institute for Molecular Science, Okazaki 444-8585, Japan}
\abst{Phonon angular momentum in chiral materials has been widely studied in spintronics and condensed matter physics. In chiral crystals, this is not the conserved quantity in contrast to the pseudo-angular momentum. To highlight this point and to understand the behavior of the angular momentum, we reexamined phonon dispersion theory based on the irreducible representation of helix and found the distinction of these angular momentum is originated from chirality.}

\begin{document}
\maketitle
Chirality is a structural property defined as a lack of inversion and reflection symmetries.
Chirality-induced phenomena in molecules and crystals have attracted immense interest~\cite{rikken2002Observation,kuwata-gonokami2005Giant,muhlbauer09Skyrmion,gohler2011Spin,inui2020ChiralityInduced}.
From this point of view,
the importance of phonon degrees of freedom has been more recognized.
As with the band dispersion of electrons and photons in crystals,
the symmetry of the crystal space group also governs the dispersion law of phonons.
The electronic band dispersion specific to crystals with helical structures has
been systematically discussed in Ref.~\citen{bozovic1984Possible}.
Although the phonon dispersion in chiral crystals has long been measured and calculated~\cite{pine1969Linear,teuchert1974Symmetry},
it is only recently that a general theory of phonon dispersion in chiral crystals
that takes into account the pseudo angular momentum (PAM) has been developed~\cite{tsunetsugu2023Theory}.
The PAM is characterized by the rotational or screw symmetry of crystals~\cite{zhang2015Chiral,zhang2022Chiral},
and plays a decisive role for selection rules
such as optical absorption represented as the exchange of PAM of phonons, which leads to the observation of the chirality-induced phonon dispersion splitting~\cite{ishito2022Truly,ishito2023Chiral}.

Our focus in this study is the angular momentum (AM) of phonons,
which is relevant to the circular motion of phonon displacement around the equilibirum posotion.
This was first introduced in elasticity theory~\cite{vonsovskii1962PhononSpin,levine1962NoteConcerningSpin},
and is recently reformulated in atomic lattice theory~\cite{zhang2014Angular,zhang2015Chiral}.
Applying this,
Ishito \textit{et al.}~\cite{ishito2022Truly,ishito2023Chiral} performed ab initio calculation of the phonon AM in chiral crystals
with emphasis on the distinction of the PAM and AM.
Based on the general theory of phonon disperion in chiral crystals~\cite{tsunetsugu2023Theory},
we develop a theory of phonon AM toward its detailed understanding of its wavenumber dependence.

We consider a single helix taken from a
Te-like crystal that belongs to the space group $P3_{1}21$ as an example.
The right-handed helix belongs to its subgroup, the line group $L3_1=\{\hat{R}^p\}_{p=0}^{3N-1}$ with $N$ being the number of unit cell,
whereas its enantiomorphic or left-handed counterpart belongs to the line group $L3_2$.
Here, $\hat{R}=(\hat{C}_{3}|c/3)$
is the screw displacement operator in Seitz notation
with $\hat{C}_{3}$ and $c/3$ being
the threefold rotation and translation along the screw axis, respectively.

The irreducible representations of the line group $L3_1$ are specified by the discrete wave number $k_n = 2n\pi/cN$ for $n = 0, \ldots, N-1$
and the rotational quantum number $m=0,\pm1$.
For the normalized phonon mode of $\bm{e}_m(k_n)$,
the PAM $m_\mathrm{PAM}$ is defined by the relation $\hat{R}\bm{e}_m(k_n)=e^{im_\mathrm{PAM}\alpha}\bm{e}_m(k_n)$ with $\alpha=2\pi/3$ being the rotation angle.
The PAM is divided into the orbital and spin part, which becomes $k_nc/3\alpha$, and $m$, respectively~\cite{ishito2022Truly}.
Thus, the spin PAM, $m$, is conserved in the phonon eigenmode.
To unveil the property of the AM in each mode,
we decompose $\bm{e}_m(k_n)$ into the modes with the definite AM as
\begin{equation}
    \bm{e}_m(k_n)
    = \sum_{s=+,-,z} a_m^s(k_n) \bm{e}_m^s(k_n)
\end{equation}
with the coefficient $a_m^s(k_n) \in \mathbb{C}$ satisfying the normalization condition, $\sum_{s=+,-,z}\lvert a_m^s(k_n)\rvert^2 = 1$.
Each mode $\bm{e}_m^s(k_n)$
is represented as the combination of $\bm{e}_\ell^s = \bm{e}_\ell \otimes \bm{e}^s$,
which is the unit vector of $\ell$th atom in the $s$th direction.
For $\bm{e}^s$, we choose the chiral basis, $\bm{e}^\pm = (\bm{e}^x\pm i\bm{e}^y)/\sqrt{2}$ and $\bm{e}^z$, because each base state have the definite AM:
$\hat{C}_3\bm{e}^s = e^{im_s\alpha}\bm{e}^s$ with $m_\pm = \pm1$ and $m_z = 0$.
Thus, the phonon AM of each mode is evaluated by
\begin{equation}
    L_m^z(k_n)
    = \hbar \sum_{s=+,-,z} m_s \lvert a_m^s(k_n)\rvert^2.
\end{equation}
This expression is proven to be equivalent to $z$-component of the AM defined in terms of the displacement and its canonical momentum vectors~\cite{zhang2015Chiral}.
In this note, we focus on the $z$-component of AM carried by phonons propagating along the chiral axis. As can be seen below, the orbital planes of the nuclei at each site tilt away from the $xy$ plane. Therefore, the AM has components in the $x$, $y$, and $z$ directions. The AM components other than the $z$-component will be presented in future papers.
By using the projection operator method,
$\bm{e}_m^s(k_n)$ is generated as
\begin{equation}
    \bm{e}_m^s(k_n)
    = \frac{1}{\sqrt{3N}} \sum_{\ell=1}^{3N}
        e^{-i(\ell-1)(k_nc/3+(m-m_s)\alpha)}
        \bm{e}_\ell^s.
    \label{eq:GFT1}
\end{equation}

Let $\bm{u}_\ell$ be the displacement of the $\ell$th atom
from its equilibrium position in the chiral basis.
In the harmonic approximation, the potential energy can be expressed as
$\Phi = \sum_{\ell=1}^{3N} (\bm{u}_{\ell+1}-\bm{u}_\ell)^\dagger \Omega^{(\ell)} (\bm{u}_{\ell+1}-\bm{u}_\ell)$,
where $\Omega^{(\ell)}$ is the force matrix.

The potential energy $\Phi$ must be totally symmetric under the symmetry operation, $\hat{R}$.
With the representation matrix for $\hat{C}_3$, $D(\hat{C}_{3}) = \diag(e^{i\alpha},e^{-i\alpha},1)$, the symmetry constraint is written as~\cite{maradudin1968SymmetryPropertiesNormal}
$\Omega^{(\ell+1)}= D(\hat{C}_{3}) \Omega^{(\ell)} D^\dagger(\hat{C}_{3})$,
which corresponds to the right-handed chirality.
The presence of the screw symmetry operation inevitably represents the chirality,
because there exists no supergroup of $L3_1$ that includes the mirror operation,
thus violating the left-handed chirality condition.
Consequently, we can obtain the chiral symmetry breaking condition,
$\Omega^{(\ell)} \ne D^\dagger(\hat{C}_{3}) \Omega^{(\ell)} D(\hat{C}_{3})$,
which is equivalent to either $\Omega_{+-}^{(\ell)}\ne 0$
or $\Omega_{+z}^{(\ell)} = \Omega_{-z}^{(\ell),\ast} \ne 0$.
Thus, the purely transverse circular or longitudinal motions do not exist in chiral crystals.

Substituting $\bm{u}_m(k_n)=u_{nm}\bm{e}_m(k_n)$,
where $u_{nm}$ is the displacement amplitude,
into the phonon potential energy $\Phi$ results in
\begin{align}
    \Phi
    = \sum_{n=0}^{N-1} \sum_{m=0,\pm1} \lvert u_{nm}\rvert^2
        \bm{a}_m^\dagger(k_n) V_m(k_n) \bm{a}_m(k_n)
    \label{eq:Phi_with_Vm}
\end{align}
with $\bm{a}_m(k_n)=(a_m^+(k_n),a_m^-(k_n),a_m^z(k_n))^\top$.
Here, $V_m(k_n)=d_m^\dagger(k_n) \Omega d_m(k_n)$
is expressed in terms of $\Omega=\Omega^{(1)}$
and $d_m(k_n)=\diag(d_m^+(k_n),d_m^-(k_n),d_m^z(k_n))$,
where each element $d_m^s(k_n)= e^{-i(k_nc/3+(m-m_{s})\alpha)}-1$
represents the relative phase difference
between the neighboring sites.
By solving the eigenvalue problem $V_m(k_n)\bm{a}_m(k_n)=\omega_m^2(k_n)\bm{a}_m(k_n)$
the square of the phonon eigenfrequencies $\omega_m^2(k_n)$ is obtained.

\begin{figure}
    \includegraphics{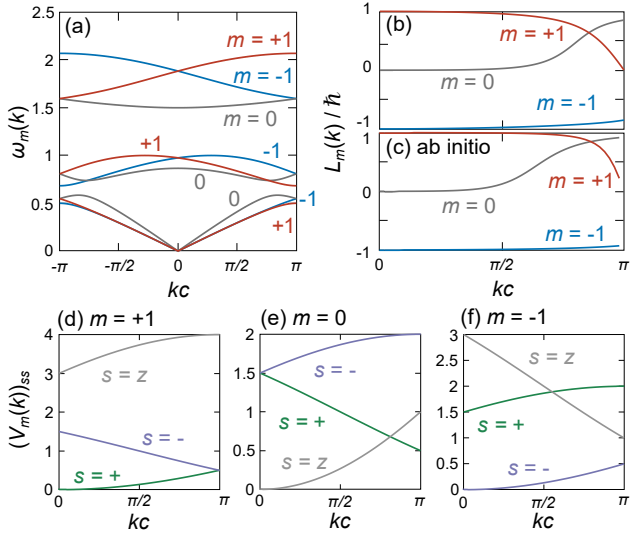}
    \caption{(Color online). (a) Phonon dispersions and (b) AMs numerically obtained for $\Omega_{xx}=\Omega_{yy}=\Omega_{zz}=\Omega_{xy}=2$ and $\Omega_{xz}=\Omega_{yz}=1$.
    (c) Phonon AMs of \ce{R-HgS} by the ab initio calculation reproduced from the source data of Ref.~\citen{ishito2022Truly}, which is licensed under a Creative Commons Attribution 4.0 International License (http://creativecommons.org/licenses/by/4.0/).
    (d)-(f) Diagonal elements of $V_m(k)$ for $m=+1,0,-1$.}
    \label{fig:result}
\end{figure}
\begin{figure}
    \includegraphics{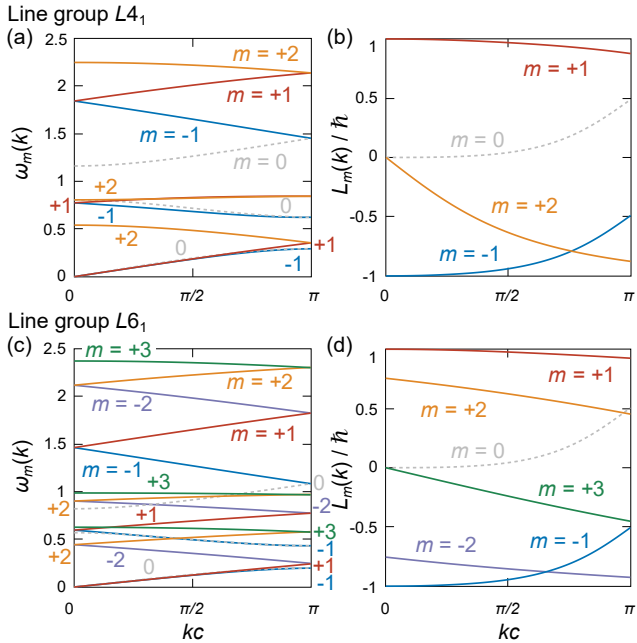}
    \caption{(Color online). (a),(c) Phonon dispersions and (b),(d) AMs of the lowest mode of each spin PAM $m$ for the helix belonging to the line group $L4_1$ and $L6_1$, respectively.}
    \label{fig:other_helix}
\end{figure}
Figure~\ref{fig:result}(a)
presents numerical results of the phonon dispersion $\omega_m(k)$,
where the splitting $\omega_{+1}(k) \ne \omega_{-1}(k)$ correctly reflects the chiral symmetry breaking.
Furthermore, the crossing $\omega_{+1}(\pi/c)=\omega_{0}(\pi/c)$ comes from the degeneracy by the time-reversal symmetry~\cite{bozovic1981Irreducible}.
In Fig.~\ref{fig:result}(b), the phonon AMs of three acoustic modes are depicted,
which is consistent with the result obtained from the ab initio calculation for \ce{R-HgS}~\cite{ishito2022Truly} (Fig.~\ref{fig:result}(c)).
The wavenumber dependence is due to the nonzero $\Omega_{+-}$ or $\Omega_{+z}$ originated from the chiral-symmetry breaking,
and clearly shows that the AM is not the conserved quantity and has no one-to-one correspondance with the PAM.
At $k=0$ and $k=\pi/c$, the time-reversal partner have the opposite AM, $L_{+1}^z(0)=-L_{-1}^z(0)$ and $L_{-1}^z(\pi/c)=-L_0^z(\pi/c)$,
and the self-conjugated states have no AM, $L_0^z(0)=L_{+1}^z(\pi/c)=0$.
The behavior of the AMs can be understood by the diagonal elements of $V_m(k)$, which are presented in Fig.~\ref{fig:result}(d)-(f).
To obtain the AMs, we need to determine each component of the vector, $a_m^s(k_n)$, which is determined as eigenvectors of the eigenvalue equation for the dynamical matrix. The $s$-state with smaller value of $(V_m(k_n))_{ss}$ gives the larger coefficient $a_m^s(k_n)$, and hence, the larger contribution to the AM.
Our approach explicitly demonstrates this tendency through the phonon wave function, Eq.~\eqref{eq:GFT1}.

The formulation presented in this study
can be straighforwardly extended to other helices.
As an example,
we cosider helices with the four and six-fold screw symmetry, which belongs to the line group $L4_1$ and $L6_1$, respectively.
The phonon dispersions and AM of these line groups are depicted in Figure~\ref{fig:other_helix},
where as in the line group $L3_1$, the chirality-induced dispersion splitting and the degeneracies at $k=0$ and $k=\pi/c$ from the time-reversal symmetry are correctly computed.

In summary,
to understand the phonon AM, 
we express the phonon dispersion theory as the eigenvalue problem of different AM modes.
The wavenumber dependence of the AM is the clear signature of the distinction between the PAM and AM, and is originated from the chiral-symmetry breaking effect.
This study will be helpful to understand the AM transfer of phonons to other quantum components, which is left for future studies.

\begin{acknowledgments}
    We thank Takuya Satoh, Hiroaki Kusunose, and Tomomi Tateishi for their valuable comments.
    This work was supported by the JSPS KAKENHI Grant No.~20J01875, and No.~21H01032,
    and Joint Research by Institute for Molecular Science (IMS program No.~23IMS1101).
\end{acknowledgments}

\bibliographystyle{jpsj}
\bibliography{25609Ref}

\end{document}